\documentstyle[11pt,paspconf]{article}

\begin{document}

\title{Cosmic Deuterium and Baryon Density}
\author{Craig J. Hogan}
\affil{Astronomy and Physics Departments, PO Box 351580, University of
Washington,
Seattle, WA 98195}

\begin{abstract}

Quasar absorption lines now  permit  a direct probe of
deuterium abundances in primordial material, with the best current
 estimate  $ (D/H)=1.9\pm 0.4 \times 10^{-4}$.
  If this is the universal primordial abundance $(D/H)_p$,
Standard Big Bang Nucleosynthesis   yields an estimate
of the mean cosmic density of baryons,
$\eta_{10}= 1.7\pm 0.2$ or $ \Omega_b h^2= 6.2\pm 0.8\times 10^{-3}$,
leading to SBBN predictions in  excellent agreement with  estimates
of primordial abundances of helium-4 and lithium-7.
Lower values of $(D/H)_p$  derived from Galactic chemical evolution models may
instead be a sign of
destruction of  deuterium and helium-3 in stars. The inferred baryon density is
compared with known baryons in stars and neutral gas; about two thirds of the
baryons are in
some still-unobserved form such as ionized gas or compact objects.
Galaxy  dynamical
mass estimates  reveal  the need for primarily nonbaryonic dark matter in
galaxy halos.  Galaxy cluster dynamics imply that the total density of this
dark matter,
while twenty or more times the baryon density, is
still well below the critical value, unless both baryons and galaxies are
concentrated in
galaxy clusters relative to the dark matter.

\end{abstract}


\section{Introduction}

The mere presence of deuterium
in the  universe confirms   the
conceptual framework of the Hot Big Bang, since
unlike other primordial nuclei  there is no other
known source. Deuterium is the unique relic  of the  Big Bang:
the total history of its cosmic production is over in only a few minutes,
followed by billions of years of slow destruction in stars.
Because it is relatively sensitive to the
baryon/photon ratio $\eta$ ($\equiv 10^{-10}\eta_{10}$), the primordial
 deuterium abundance
is also the best way of measuring
the amount of matter in the universe.

 Until
recently, deuterium was only measured within our highly chemically evolved
Galaxy,
  which has  destroyed a large but uncertain fraction of its initial deuterium.
Measurements of   deuterium abundance in QSO absorbers
now allow a much more direct   estimate of primordial abundance, in pristine
material that
has suffered little stellar processing.
The Big Bang
prediction is approximately fitted by
$$
(D/H)_p\approx 4.6\times 10^{-4}\eta_{10}^{-5/3}
= 1.9\times 10^{-4} ({\eta_{10}/1.70})^{-5/3},
$$
so an estimate  of primordial D/H with an   absolute accuracy  of $\pm 20\%$
yields an estimate of   primordial  baryon/photon ratio accurate to $\pm 12\%$,
a very
good precision by the standards of   cosmic bookkeeping.\footnote{The accuracy
of the fitting
formula is about 7\%  in this range, which is almost as good as the estimated
theoretical
error (Krauss and Kernan 1995). Note that the
conversion to a physical density also requires a knowledge of CBR temperature:
$\Omega_b h^2= 3.631\times 10^{-3}\eta_{10}T_{2.726}^3$.}
 This
accuracy is   attainable from measurements in QSO absorption line systems.
I survey here the strengths and limitations of the QSO technique,
compare early QSO results with Galactic estimates of $(D/H)_p$, and summarize
cosmological
implications of  the low  value of baryon density estimated from the high
$(D/H)_p$.

\section{Deuterium in Quasar Absorbers}

Certainly in the long run, and I argue here even at the present,
the best way to measure primordial D/H is through absorption
lines of distant quasars. The principles are discussed by Jenkins in this
volume.
Profiles of the Lyman series lines of HI and DI and the Lyman limit optical
depth
combine to give good absolute column densities for both species in many
situations, which
in principle
yield an absolute abundance almost free of ionization, temperature or density
corrections in
atomic gas. Accurate absolute column density information comes from optically
thin or damped
absorption lines of each species, and from the Lyman limit; ionization and
recombination
processes are nearly identical for both, so little detailed physical or
geometrical
modeling is required.
 Metal
lines from the same clouds provide an independent measure of the amount of
chemical
processing, and in some cases the metal abundances are so small that
significant
deuterium depletion
can essentially be ruled out. Eventually, many clouds will
 be measured along different lines
of sight, testing the universality of the abundance in different places all
over our past
light cone, confirming (or not) that the past worldlines of matter were similar
over an
enormous spacetime volume, testing both the large scale homogeneity of
spacetime (the
``Cosmological  Principle'') and the small scale homogeneity of matter.

The greatest weakness with the QSO technique is the ambiguity caused by
the unknown distribution of material in velocity, or the  problem of
interlopers--- clouds of hydrogen gas which happen by chance to lie at the
redshift expected
for   deuterium associated with some other hydrogen cloud.  Nature does not
provide  us
with a   system cleanly arranged in velocity, so any detection of deuterium
must be regarded
as suspect until this possibility is  dealt with.

The first way to deal with interlopers is by seeking those relatively
 rare narrow-line systems where  the turbulent component of linewidth is small
and thermal
width itself is also small, close to the minimum temperature consistent with
photoionization
heating. Because deuterium is heavier, the Doppler parameter
$b \equiv \sqrt{{2kT}\over{m}}$ is smaller; in a purely thermally broadened
profile,
$b= 13 T_{4}^{1/2}$ for hydrogen, but only
$9 T_{4}^{1/2}$ for deuterium. Gas under equilibrium
conditions in these Lyman Limit absorbers is not expected to be
cooler than $10^{4}$ K (usually, $T_{4} =
T/10^{4}K \approx 1 \ to \ 3$, over a very wide range for the ionization
parameter; see
Donahue and Shull, 1991), so   lines with $b\le 10$ are
unlikely to be hydrogen interlopers.  The ratio of the fitted
Doppler parameter  for D candidates to that of their H counterparts is also an
important
clue. Turbulent broadening tends to give both lines the same profile (so
different shapes
are not
always expected), but a situation where the deuterium line is narrower than the
hydrogen by a factor between   $1/\sqrt{2}$ and 1 is again unlikely to be an
interloper.
Rugers and Hogan (1995) argue that the
candidate deuterium absorber in Q0014+813 in fact displays both of these
signatures of real deuterium, and so
is likely to be a real measurement of D/H.
  It  yields an abundance estimate
$D/H= 1.9\pm 0.4\times 10^{-4}$ (Songaila et al 1994, Carswell et al 1994,
Rugers and Hogan 1996 and this volume).

The errors in this estimate are real measured errors, in the sense that they
reflect the
 total uncertainty in the fitted column densities, including ambiguities in
Doppler
parameter and velocity. They do not include systematic ``model
errors'', the most extreme example being a hydrogen interloper, and for this
reason
it is risky to rely on only one  example, however clean. Also, one should bear
in mind that
although these are true ``$1\sigma$'' statistical uncertainties, the
distribution of allowed
values is highly nongaussian,  so we do not have a good estimate of the
probability of  larger $2\sigma$ excursions from the fitted values. The ranges
quoted
throughout must be taken only as current best guesses, with the possibility of
large
departures not well constrained. This is main reason why more clean absorbers
are urgently required.

Unfortunately   clean conditions are not the rule. One can invoke
statistical arguments based on larger samples of
less trustworthy  candidates--- is there a statistical
excess of lines at
82 km/sec to the blue of hydrogen (either relative to the red or relative to
adjacent velocities)? Is there a statistical tendency of
fits to improve with deuterium? There are several examples\footnote{One
possible
counterexample with $D/H\approx 2\times 10^{-5}$ may have been found by Tytler
and Fan (1994)
on the line of sight to  Q1937-1009. Carswell (1995) quotes a lower limit in a
system in
Q0420-388 of $D/H\ge 2\times 10^{-5}$, with a  best guess of
$2\times 10^{-4}$. Songaila and Cowie have a good fit for $2\times 10^{-4}$
in Q0956+122. See the contribution by Rugers and Hogan in this volume for
another example
in GC0636+68, which yields $\log (D/H)=-3.95 \pm 0.54.$}
 where
this is the case, but
 so far the samples are not large enough to make statistical
tests meaningful, although the dozen or so absorbers
in two spectra we have studied so far
indicate    that
that   within the large errors the data are consistent with a universal
abundance
of the order of $10^{-4}$. The most trustworthy system to make a   measurement
remains the   pair of absorbers in Q0014+813.

We now make a major leap of reasoning, and take the Q0014 measurement as an
estimate
of the primordial abundance. The justifications for this leap are (1) The Big
Bang is the
only known source of this deuterium, so any measurement is a reasonable lower
limit on
$(D/H)_p$; (2) The Q0014 cloud is extremely metal-poor and is unlikely to have
cycled an
appreciable faction of its
material through stars (that is, $D$ destruction by stellar cycling should be
accompanied by noticeable enrichment), so it gives a reasonable upper limit on
$(D/H)_p$; (3)
Although it is only one site, it is still the best one we have at this
writing--- the most
accurately measured  and most pristine.  For this value of $(D/H)_p$,
Big Bang nucleosynthesis implies $\eta_{10}= 1.7\pm 0.2$, which gives excellent
concordance of Big Bang predictions with with the most straightforward
interpretations
of helium-4 and lithium-7 abundance data (Copi et al 1995ab, Hata et al 1995,
and  Schramm and Steigman, this volume.)  This gives us the confidence
to trust in the Big Bang picture enough to use $D/H$ as a probe of cosmic
baryon density.

\section{Deuterium in Galactic Chemical Evolution}

The value $(D/H)_p= 2.0\pm 0.4\times 10^{-4}$ is surprisingly high.
Smaller values have previously been quoted as {\it upper} limits on $(D/H)_p$,
giving {\it lower} limits on $\Omega_b h^2$, based on abundances in our Galaxy,
together with the assumption that the sum $(D+{^3He})/H$ cannot decrease.

In the Galaxy today, the deuterium abundance is less than
$ 2\times 10^{-5}$ (Linsky et al 1993, 1995, and Lemoine, this volume).  This
can be
understood if the interstellar gas has almost all been processed through at
least the outer
envelopes of stars. A high primordial  value $(D/H)_p\ge 10^{-4}$ requires that
stellar
processing in the Galaxy
destroys  not only
deuterium but also its principal burning product, helium-3, in order to agree
with the
low values of this isotope found for the presolar material: analysis of the
solar wind and
meteorites reveals that the presolar nebula  had about
$^3He/H\approx 1.5\pm 0.5\times 10^{-5}$,
$D/H\approx 2.7\pm 2\times 10^{-5}$, and
$(D+^3He)/H\approx 4\pm 2\times 10^{-5}$ (Copi et al 1995ab, Hata et al 1995).
 It is probably
necessary to have some helium-3 destructive
 mechanism
also in order to explain the interstellar observations of helium-3, which are
highly variable and sometimes very low,
of the order of $0.6- \  6\times 10^{-5}$ (Balser et al 1994; Wilson and Rood
1994).

While deuterium is destroyed by stars even before they enter the main sequence,
and there is no doubt that helium-4 increases  due to stellar processing,
helium-3  is
both created and destroyed by stars. It is fairly abundant in the interiors of
 main sequence stars in the temperature range of H burning, but is destroyed at
higher
temperatures. The helium-3 ejected back into the ISM when a star dies depends
in detail on
what the material of the star does after it leaves the main sequence
and before it throws off its hydrogen envelope. Although the nuclear
physics is well understood,  details of the movement of material between
different
temperatures, and how it is ejected into the interstellar medium, are murky.

In galactic chemical evolution models, low mass stars (one or two solar masses)
dominate
stellar cycling and
destroy the bulk of the pregalactic deuterium, so it is these stars that must
 destroy the
helium-3 (Galli et al 1995, Olive et al 1996).   There is a way this could
happen: if the
material in the stellar envelope on the giant branch is brought to high
temperature
($1.5\times 10^7 $K or so) before it is ejected, the helium-3 would be
destroyed. There is
evidence for just such a   process  (``Giant Branch Mixing'',  ``Cool
Bottom Processing'' or ``extra-mixing''; Hogan  1995, Wasserburg et al  1995,
Charbonnel
1995)--- from the change in C and N isotopic composition as stars leave the
main sequence,
from O isotope mixtures in meteorites,  and from the continued decrease in
lithium after cool
giants leave the main sequence. There is thus at least a plausible way the
helium-3 could be
destroyed. Charbonnel (1995) has recently shown a detailed model which destroys
helium-3
between 1 and 2 solar masses, but not above 2--- thereby plausibly explaining
the above
facts. The three nearby planetary nebulae which have very high helium-3
(indeed, so high
that they must be atypical; Wilson   and Rood 1992) could simply have
 progenitor masses
above 2 solar masses. Although we do not have a clear constraint on   the
integrated  stellar population ejecta, it is clearly not safe to assume that
the Galaxy
cannot reduce its helium-3 abundance with time. This is why the QSO
measurements, even in
their present unsettled state, are a more reliable measure of primordial $D/H$
than
extrapolating from present Galactic values.

\section{Cosmic Baryon Inventory}

The   inventory of things which must be made of baryons includes HI absorbers
($\Omega h\ge
0.003$, Wolfe 1993), galactic stars ($\Omega\approx 0.002$), and cluster gas
($\Omega
h^{0.5}\approx 0.001$) (Persic   and Salucci 1992), with errors on these
quantities typically
$\pm 30\%$.  Elliptical and spiral
galaxies contribute about equally to  the stellar mass, although spirals are
more numerous
 and with their younger populations
dominate the blue and visible  light by about a factor of two (Schechter and
Dressler 1987).
In addition there is
  an ionized gas component, including for example the Ly$\alpha$ forest clouds,
whose density
could be larger than  all of these or smaller than any of them.
The HI gas
density is only this large at high redshift (that is, it might convert into
stars at low
redshift),  so the minimum requirement is to  count  things at only one
redshift. Taking
$h=0.75$,  it could be that all known baryonic things could be accounted for
with as little as
$\Omega_b\approx 0.004$. The deuterium estimate
of $\Omega_b=0.011$
thus provides amply  sufficient baryons  to make all the things that need to be
made of
baryons. There must still  be some unaccounted baryons,probably in the form of
ionized gas
and/or compact objects (Carr 1994).
The baryon to galaxy mass ratio is about 5,  and the baryon to (galaxy+known
gas)
ratio is about 3. Thus
 within the errors,  the  uncounted baryons  could comprise up to, but
probably not more than about twice that already seen. This is quite different
from
the
previous situation, where lower estimates of $(D/H)_p$ required more than 90\%
of the
baryons  to be dark or in an ionized IGM. It leads to a tidy model of galaxy
formation, which accounts for most of the facts about galaxies and QSO
absorbers from
$z=0$ to $z=3$, where baryons reside for the most part in
gas, stars and compact objects in the vicinity of galaxies today (Fukugita et
al 1996).  The
abundance of MACHOs in the Galactic halo (Alcock et al. 1995) is consistent
with such a low
density of baryons.

\section{Nonbaryonic Dark Matter in Galaxies}

On the other hand the  estimated   baryon density is not enough to account for
the known dark matter in the universe. This problem is well known,
although   the need for
nonbaryonic dark matter is greater than before, and now extends even to
galactic halos.

 Using
the observed integrated blue luminosity density of galaxies,
$L_B=1.93^{+0.8}_{-0.6}
\times 10^8 h$   solar  units per Mpc$^3$
(Efstathiou et al. 1988), we can write the
physical  baryon density estimate as a
 global baryon mass density to blue luminosity density ratio,
$\rho_b/L_B\approx 9 h^{-1}$ in solar units.
For our Galaxy, the mass to blue light ratio is about 60 if the mass
is $10^{12}$ solar masses(Binney and Tremaine 1987), a typical dynamical
estimate (Zaritsky
et al 1993, Peebles 1995).
The mass to light ratio from galaxy rotation curves is
often inferred in typical spirals---
the same types that dominate the luminosity density---
to be at least $30h^{1}$ (Rubin 1993).
If this ratio is
universal for galaxies,\footnote{In  some gas-rich, star-poor dwarfs the
rotation curve
 can be measured out to many scale lengths; there are
 cases (eg, DDO 154) where $M/L$ is more than $50h^{1}$ (Rubin 1993). Since
these are not
 typical galaxies (the luminosity per baryon is known to be less than usual),
it is
 hazardous to use their
 $M/L$ as a universal value (their bias is to overestimate $\Omega$).}
 the global density of galactic dark matter is $\Omega\ge 0.02$.

Galactic dark matter is thus probably not made mostly of baryons--- a stronger
statement about nonbaryonic dark matter than was possible with the earlier
higher baryon
density estimates.

\section{Global Dark Matter Density}

 The need for nonbaryonic dark matter is much greater if rich
galaxy clusters, with dynamical mass to galaxy  light ratios of about
300$h^{1}$, fairly
represent the global M/L, implying $\Omega\approx 0.2$ (Binney and Tremaine
1987). It is
possible however that galaxy formation is more or less efficient in clusters
than in the
field, so that their M/L is not the universal one. The appeal to greater
efficiency is
necessary  especially in models with an overall dark matter density close to
$\Omega=1$.

Physical models of this ``biasing'' are  constrained by the fact that the
baryon to galaxy
ratio in clusters, as measured directly, is close to its global value, as
inferred from
nucleosynthesis.  White et al. (1993) estimate for example that the Coma
 cluster\footnote{Although they used  Coma specifically, similar numbers are
obtained in other
clusters, using other techniques; for example, Squires et al.1995 derive from
weak lensing
mass estimators an upper limit of
$M_{gas}/M\le (0.04\pm 0.02)h^{-3/2}$ and $M/L_B=(440\pm 80 )h$ for the inner
400kpc of
A2218}
 has a ratio of baryons to dark matter within the
virial radius of
$(0.05
\pm 0.01)h^{-1.5}$ in the form of gas, and $0.009\pm 0.002$ in the form of
stars;
  the higher baryon to galaxy mass ratio of 8 (for $h=.75$, compared to the
above
global estimate
of 5) indicates if anything that galaxy formation was less efficient there than
in the
field, unless the gas mass is overestimated.
This point does not depend on
estimates of cluster mass. This  argues against classical biasing (i.e.,
protocluster
galaxies forming earlier and more efficiently than protofield galaxies) as a
way
of reconciling
cluster M/L with $\Omega=1$: the only way to much higher cosmic density is to
increase the
ratio of baryons to dark matter in clusters relative to the cosmic mean, by
about a factor
of five.

White et al. dispense with the galaxies altogether, and use the
baryon to dark matter ratio directly.
They
use simulations and physical models of cluster collapse to show that
composition within the virial radius ought to be
representative of the global baryon/dark matter mix for collisionless cold dark
matter
models.
 The above
estimate  for
$\Omega_bh^2$ yields $\Omega= 0.12 h^{-0.5}$---
again implying an open universe, or else a flat universe with an
unclustered mass density, such as a cosmological constant.

 \acknowledgments
This work was supported at the University of Washington by NSF grant AST
9320045 and
NASA grant NAG-5-2793.

\end{document}